\documentstyle[preprint,aps,prd]{revtex}
\draft
\begin{document}
\title{Ground-State Properties of Magnetically Trapped Bose-Condensed
Rubidium Gas}

\author{Gordon Baym$^1$ and Christopher Pethick$^{1,2,3}$}

\address{$^1$Department of Physics, University of Illinois at
        Urbana-Champaign, 1110 W. Green St., Urbana, IL 61801 \\
$^2$Nordita, Blegdamsvej 17, DK-2100 Copenhagen \O, Denmark,\\
$^3$Institute for Nuclear Theory, University of Washington,
 Box 351550, Seattle, WA 98195\\}

%\date{} 24 August 1995

\maketitle

\begin{abstract}

    We give a quantitative account of the ground-state properties of the
dilute magnetically trapped $^{87}$Rb gas recently cooled and Bose-Einstein
condensed at nanokelvin-scale temperatures.  Using simple scaling arguments,
we show that at large particle number the kinetic energy is a small
perturbation, and find a spatial structure of the cloud of atoms and its
momentum distribution dependent in an essential way on particle interactions.
We also estimate the superfluid coherence length and the critical angular
velocity at which vortex lines become energetically favorable.\\

\end{abstract}


    In a remarkable experiment, Anderson et al.  \cite{nist} have cooled
magnetically trapped $^{87}$Rb gas to nanokelvin-range temperatures, and
observed a rapid narrowing of the velocity distribution and density profile,
which is interpreted as the onset of Bose-Einstein condensation.  Trapped atom
clouds are new systems, beyond liquid $^4$He \cite{london} and excitons in
semiconductors \cite{wolfe}, in which particles obeying Bose-Einstein
statistics condense at low temperatures \cite{BEC93,huletbec}; indeed, the
condensation of trapped atoms has been a ``Holy Grail" of atomic physics
\cite{burnett}.  In this paper we give simple quantitative arguments, taking
into account effects of the repulsive interatomic interactions, to determine
the properties of the quantum ground state of the trapped $^{87}$Rb system,
including its geometry, momentum distribution, coherence length and critical
angular velocity for vortex formation.

    In the experiment, the gas is magnetically trapped in an effective
three-dimensional harmonic well cylindrically symmetric about the z-axis, with
tunable angular frequencies $\omega^0_z$ in the axial (z) direction and
$\omega^0_\perp = \omega^0_z/\sqrt{8}$ in the transverse (xy) plane.  The
oscillators are characterized by lengths $a_\perp =
(\hbar/m\omega^0_\perp)^{1/2}$ and $a_z = (\hbar/m\omega^0_z)^{1/2}$, where
$m$ is the atomic mass.  During the condensation phase, with $\omega^0_z/2\pi
= 208$ Hz, and $a_\perp \approx 1.25\times10^{-4}$ cm, the distribution
rapidly sharpens with falling temperature, as a macroscopic number, $N_0$, of
the Rb atoms begin to occupy the lowest mode of the well.  In the absence of
interparticle interactions the lowest single-particle state has the familiar
wave function,
\begin{eqnarray}
\phi_0({\vec r}) = \frac {1}{\pi^{3/4}a_\perp a_z^{1/2}}
     e^{-m(\omega^0_\perp r_\perp^2 + \omega^0_z z^2)/2\hbar},
\label{ground}
\end{eqnarray}
where $\vec r_\perp$ is the component of $\vec r\,$ in the xy-plane.  The
density distribution, $\rho({\vec r\,}) = N_0 \phi_0({\vec r})^2$, is
Gaussian.  However, interatomic interactions strongly modify the particle
structure in the well.

    The low-energy interactions between polarized $^{87}$Rb atoms are
repulsive, and are described by an s-wave triplet-spin scattering length, $a$,
determined to be in the range $85a_0 < a < 140a_0$, where $a_0$ is the Bohr
radius \cite{gardner}.  In the limit in which the density varies slowly on a
scale $a$, the interaction energy of the gas per unit volume is given by
$E_{int} = (2\pi\hbar^2 a/m)|\rho({\vec r\,})|^2$.  The repulsive interactions
favor a reduction of the density from the free particle situation, $N_0
\phi_0({\vec r})^2$.  As the number of particles increases, the first effect
of interactions is to cause the cloud of particles to expand in the transverse
direction, where the restoring forces are weaker.  With further increase in
the number, the cloud expands in the z-direction.  The eventual size of the
cloud is determined, in the limit in which the interparticle interactions
dominate, by a balance between the harmonic oscillator and interaction
energies.

    To see the physics of the balance, let us neglect the anisotropy of the
oscillator potential and assume that the cloud occupies a region of radius
$\sim R$, so that $\rho \sim N/ R^3$; then the scale of the harmonic
oscillator energy per particle is $\sim m\omega_\perp^2 R^2/2$, while each
particle experiences an interaction energy with the other particles $\sim
(4\pi\hbar^2 a/m)(N/R^3)$.  The characteristic length scale is thus $\sim
a_\perp \zeta$, where the dimensionless parameter characterizing the system is
\begin{eqnarray}
 \zeta \equiv (8\pi N a/a_\perp)^{1/5}
\approx 4.21 \left[(a/100a_0) (N/10^4)(10^{-4}{\rm cm}/a_\perp)\right]^{1/5};
\label{zeta}
\end{eqnarray}
under the conditions of the experimental trap with large $N$, $\zeta \gg
1$.  The kinetic energy, on the other hand, is of order $N\hbar^2/(2mR^2)$, so
that the ratio of the kinetic to interaction or oscillator energies is of
order $\zeta^{-4} \sim N^{-4/5}$; for $N\sim 2000$, the kinetic energy in the
condensed phase is a factor $\sim 200$ smaller than the interaction energies.

    To estimate the interaction effects more quantitatively we examine the
ground state of the system in terms of its order parameter $\psi({\vec r\,})$,
where $\int d^3 r |\psi({\vec r\,})|^2 = N$.  [We do not distinguish $N$ and
$N_0$ in the weakly interacting system at zero temperature.] In the Hartree
approximation, in which $\psi({\vec r\,})/N^{1/2}$ is the lowest single
particle mode, the ground state energy of the system in given by a
Ginzburg-Pitaevskii-Gross energy functional \cite{gross},
\begin{eqnarray}
E(\psi) = \int d^3 r \left[ \frac{\hbar^2}{2m}|\nabla \psi(\vec r\,)|^2
                  +\frac{m}{2}\left((\omega^0_\perp)^2 r_\perp^2
                   + (\omega^0_z)^2 z^2\right)|\psi({\vec r\,})|^2
           +\frac{2\pi\hbar^2 a}{m}|\psi({\vec r\,})|^4\right].
\label{gp}
\end{eqnarray}
This approach is familiar in prior studies of Bose-condensed polarized
atomic hydrogen \cite{tony,siggia}; see also Refs.
\cite{hulet,ruprecht,kagan}.  The Rb experiments, with lower density, larger
atomic mass and stronger interactions, fall, however, in a rather different
parameter range.

    For a first solution we take $\psi$ in the form of the ground-state
wave-function, Eq.  (\ref{ground}):
\begin{eqnarray}
  \psi({\vec r\,}) =
     \frac{N^{1/2}}{\omega_\perp^{1/2}\omega_z^{1/4}}
      \left(\frac{m}{\pi\hbar}\right)^{3/4}
      e^{-m(\omega_\perp r_\perp^2 + \omega_z z^2)/2\hbar},
\label{var}
\end{eqnarray}
with effective frequencies, $\omega_\perp$ and $\omega_z$, treated as
variational parameters.  Substitution of (\ref{var}) into (\ref{gp}) yields
the ground state energy
\begin{eqnarray}
E(\omega_\perp,\omega_z) = N\hbar\left(
     \frac{\omega_\perp}{2}  +\frac{(\omega^0_\perp)^2}{2 \omega_\perp}
     +\frac{\omega_z}{4}  +\frac{(\omega^0_z)^2}{4 \omega_z}
     +\frac{N am^{1/2}}{(2\pi\hbar)^{1/2}}\omega_\perp\omega_z^{1/2}\right);
\label{gndstate}
\end{eqnarray}
minimizing $E$ with respect to $\omega_\perp$ we derive
\begin{eqnarray}
   \omega_\perp = \omega_\perp^0/\Delta,
\label{effoscperp}
\end{eqnarray}
where
\begin{eqnarray}
  \Delta = \left(1+\frac{\zeta^5}{(32\pi^3)^{1/2}}
  \left(\frac{\omega_z}{\omega_\perp^0}\right)^{1/2}\right)^{1/2}.
\label{delta}
\end{eqnarray}
Interactions, by reducing the effective transverse oscillator frequency by
$\Delta$, spread out the distribution in the transverse direction by a factor
$\Delta^{1/2}$; when $N$ is sufficiently large that $\zeta \gg 1$,
$\Delta^{1/2}\approx 2.55 \left[(N/10^4)(a/100a_0)(10^4{\rm cm}/a_\perp)
\right]^{1/4} (\omega_z/\omega_\perp^0)^{1/8}$.

    Spreading in the z-direction begins to become significant when the
interaction energy per particle becomes comparable with $\hbar \omega_z$;
using (\ref{effoscperp}) and minimizing the resultant ground state energy,
$E(\omega_z) = N\hbar\left(\omega_\perp^0 \Delta +\omega_z/4
+(\omega^0_z)^2/(4 \omega_z)\right)$, with respect to $\omega_z$, we see that
this condition is $(Na/a_\perp) \raisebox{-.5ex}{$\stackrel{>}{\sim}$}
(\omega_z^0/\omega_\perp^0)^{1/2}$, which is realized under the experimental
conditions.  In the limit $\zeta \gg 1$, the kinetic energy terms in
(\ref{gndstate}) are negligible, and we find the shift in the frequency in
the z-direction,
\begin{eqnarray}
  \frac{\omega_z}{\omega_z^0} = \frac{2(\pi\lambda)^{3/5}}{\zeta^2},
\label{zshift}
\end{eqnarray}
where $\lambda \equiv \omega_z^0/\omega_\perp^0$; the leading contribution
to the energy per particle is
\begin{eqnarray}
 \frac{E}{N}=\frac{5\zeta^2}{8\pi^{3/5}} \lambda^{2/5} \hbar\omega_\perp^0
   \propto N^{2/5}.
\label{Eosc}
\end{eqnarray}

    To obtain the ground state wave function more precisely, we minimize the
total energy (\ref{gp}) with respect to $\psi$, keeping the total number of
particles fixed, and thus derive the non-linear Schr\"{o}dinger equation
\begin{eqnarray}
   \left[
   -\frac{\hbar^2}{2m} \nabla^2
   + \frac{1}{2}m(\omega_\perp^0)^2( r_\perp^2 +\lambda^2 z^2)
    +\frac{4\pi \hbar^2 a}{m}|\psi(\vec r\,)|^2\right]\psi(\vec r\,)
     =\mu \psi(\vec r\,),
\label{schr}
\end{eqnarray}
where $\mu$ is the chemical potential.  The physical scales are
conveniently brought out by rescaling the lengths, letting $\vec r\, =
a_\perp\zeta \vec r_1$, and writing $\psi(\vec r\,) = (N/\zeta^3
a_\perp^3)^{1/2}f(\vec r_1)$, where $\int d^3 r_1 |f|^2 = 1$; then
(\ref{schr}) becomes the dimensionless equation,
\begin{eqnarray}
   \left[-\frac{1}{\zeta^4}\nabla_1^2
   + r_{1\perp}^2 + \lambda^2 z_1^2  + |f(\vec r_1)|^2\right] f(\vec r_1),
    = \nu^2 f(\vec r_1),
\label{schr1}
\end{eqnarray}
where $\nu^2 \equiv (2\mu/\zeta^2\hbar\omega_\perp^0)$.

    In the limit of large $N$, we can obtain an essentially exact expression
for the ground state wave function, corresponding to the Thomas-Fermi
approximation, by neglecting the kinetic energy term, which falls as
$\zeta^{-4}$; then
\begin{eqnarray}
  f(\vec r_1)^2 = \nu^2 - r_{1\perp}^2 - (\lambda z_1)^2
 \label{f}
\end{eqnarray}
in the region where the right side is positive, and $f$ =0 outside this
region.  This form for the wave function is good, except where the density is
small, in which case the kinetic energy causes the wave function to vanish
smoothly.  The normalization condition on $f$ implies that $\nu =
(15\lambda/8\pi)^{1/5}$, which translates into the relation between $\mu$ and
$N$:
\begin{eqnarray}
   \mu = \frac{\hbar \omega_\perp}{2}
    \left(\frac{15\lambda}{8\pi}\right)^{2/5}\zeta^2
     = \frac{\hbar \omega_\perp}{2}
    \left(\frac{15\lambda Na}{a_\perp}\right)^{2/5}.
\label{mu}
\end{eqnarray}
Since $\mu=dE/dN$, the energy per particle is simply $E/N =(5/7)\mu$, a
result smaller than the effective oscillator frequency calculation
(\ref{Eosc}) by a factor $(3600\pi)^{1/5}/7 \approx 0.92$.  The central
density of the blob is $\rho(0) = m\mu/4\pi\hbar^2 a$.

    In the limit of large $N$, the transverse radius of the cloud is given by
$R/a_\perp= (15\lambda Na/a_\perp)^{1/5}$, and the half-height in the
z-direction is $Z= \lambda R$.  For $N$ = 2000, $a = 100 a_0$, and
$\omega^0_\perp /2\pi =208/\sqrt8$ Hz, one has $R/a_\perp \approx 3.24$.  When
the oscillator ``spring constants" are relaxed in the trap by a factor 75, the
equilibrium configuration of the cloud then has $R/a_\perp \approx 2.61$.
These increases are consistent with observations \cite{nist} made when the
cloud is finally released from the trap \cite{holland}.  We also note that for
large $N$ the aspect ratio $R/Z$ equals $\lambda$, whereas in the absence of
interactions it is $\lambda^{1/2}$; thus for the experimental conditions, one
would expect the aspect ratio to be $\sqrt{8}$.

    For the Thomas-Fermi wave function, the momentum distribution is
proportional to $|J_2(\kappa)/\kappa^2|^2$, where $J_2$ is the Bessel function
of order 2, $\kappa^2 = (\zeta a_\perp)^2\left(p_\perp^2
+(p_z/\lambda)^2\right)$, and $\vec p$ is the particle momentum.  When
$Na/a_\perp >> 1$, the width of the momentum distribution is much less than
that for a single particle in the oscillator potential.

    The sound velocity, $c_s$, in the interior of the cloud is given by $c_s^2
= (\rho/m)(\partial\mu/\partial \rho) = \mu/m$, which in the large $N$
limit equals $(\hbar\omega_\perp^0)(15\lambda Na/a_\perp)^{2/5}$.  In this
limit, the lowest mode of excitation in the transverse direction of the system
has frequency of order $R/c_s \sim \omega_\perp^0$.

    The superfluid coherence length, $\xi$, which determines the distance over
which the condensate wave function can heal, can be estimated by equating the
kinetic energy term in Eq. ({\ref{schr}}), $\sim \hbar^2/(2m\xi^2)$, to the
interaction energy, which yields
\begin{eqnarray}
  \xi^2 =(8\pi \rho a)^{-1},
\label{coherence}
\end{eqnarray}
where $\rho$ is the local density.  With the central density of the cloud
computed in the Thomas-Fermi approximation, $\rho(0)=m \mu/(4 \pi \hbar^2 a)$,
we have
\begin{eqnarray}
 \frac{\xi}{R}= \left(\frac{a_\perp}{R}\right)^2=
 \left(\frac{a_\perp}{15\lambda Na}\right)^{4/5}.
\end{eqnarray}
Thus when the number of particles is sufficiently large that the
Thomas-Fermi approximation is valid, the coherence length is small compared
with the size of the blob, and the system should exhibit superfluid properties
more like those of a bulk superfluid than an atomic nucleus, where $\xi \sim
R$.

    An experimentally important confirmation of Bose-Einstein condensation
would be the observation of formation of a vortex line in a rotating system.
The critical angular frequency, $\Omega_{c1}$, at which it becomes
energetically favorable for a vortex line to be created under rotation about
the z-axis is
\begin{eqnarray}
    \Omega_{c1}\sim \frac{\hbar}{mR^2}ln(R/\xi).
\end{eqnarray}
For cloud radii $\sim 5\times$ 10$^{-4}$cm,  this value corresponds to a
rotation frequency of order 10 Hz.

    Finally we consider the case of atoms, such as spin-polarized $^{85}$Rb
\cite{gardner} or $^7$Li \cite{hulet}, with a negative scattering length,
corresponding to a low energy attractive interaction.  A uniform state of such
atoms at low density would be unstable to formation of long-wavelength density
waves, signalling a gas-liquid phase transition.  However, as discussed
theoretically in \cite{ruprecht,kagan}, and seen experimentally in
\cite{huletbec}, the physics in a trap is different; this can be understood in
the present context by considering the variational calculation above.  With
increasing $N$, the spatial extent of the wave function is reduced.  Provided
$\Delta^2 =
\left(1-(2/\pi)^{1/2}(N|a|/a_\perp)(\omega_z/\omega_\perp^0)^{1/2}\right)$
(cf.  (\ref{delta})) remains positive, the kinetic energy term is able to
stabilize the system.  However, if $\Delta^2$ becomes negative, the attractive
forces overwhelm the kinetic energy, and the cloud becomes unstable to
collapse.  The critical number of particles for collapse is $\sim
(\pi/2\lambda)^{1/2} a_\perp/|a|$.  In $^{85}$Ru, for which $-1000 a_0 < a <
-60 a_0$ \cite{gardner}, under the experimental conditions in Ref.
\cite{nist} with $\omega_z/2\pi$ = 208 Hz, this number is $\sim 20-300$; in
the $^7$Li trap of Ref.  \cite{hulet} it is $\sim 3000$.  The final state of
the collapsed cloud is determined by the shorter-range repulsive components of
the interatomic potential.

    To summarize, our calculations provide quantitative results that confirm
and extend the qualitative considerations in Ref.\cite{nist} on the effect of
particle interactions on the properties of a cloud of bosons.  Experimental
confirmation of the dependence on trap parameters, particle number and atomic
properties of the size of the cloud and the momentum distribution would give
one increased confidence in the interpretation of the data.

    We are grateful for the hospitality of the Aspen Center for Physics, where
this work was largely carried out.  In Aspen we had useful conversations with
Andrei Ruckenstein on this topic.  We are grateful to Eric Cornell for very
helpful instruction on the experiments, and to D. G. Ravenhall for helpful
advice.  This work was supported in part by NSF Grants NSF PHY94-21309 and NSF
AST93-15133, and NASA Grant NAGW-1583 .

\end{document}